\documentclass[epj]{svjour}
\usepackage{graphicx}
\begin{document}
\title{Breaking SU(3) spectral degeneracies in heavy deformed nuclei}
\author{Dennis Bonatsos\inst{1}\fnmsep\thanks{\email{bonat@inp.demokritos.gr}} \and \underline{I.E. Assimakis}\inst{1}\and Andriana Martinou\inst{1}  \and S. Peroulis\inst{1}\and S. Sarantopoulou\inst{1}  \and N. Minkov\inst{2} }
\institute{Institute of Nuclear and Particle Physics, National Centre for Scientific Research 
``Demokritos'', GR-15310 Aghia Paraskevi, Attiki, Greece
 \and Institute of Nuclear Research and Nuclear Energy, Bulgarian Academy of Sciences, 72 Tzarigrad Road, 1784 Sofia, Bulgaria}
\abstract{Symmetries are manifested in nature through degeneracies in the spectra of physical systems. In the case of heavy deformed nuclei, when described in the framework of the Interacting Boson Model, within which correlated proton (neutron)  pairs are approximated as bosons, the ground state band has no symmetry partner, while the degeneracy between the first excited beta and gamma bands is broken through the use of three-body and/or four-body terms. In the framework of the proxy-SU(3) model, in which an approximate SU(3) symmetry of fermions is present, the same three-body and/or four-body operators are used for breaking the degeneracy between the ground state band and the first excited gamma band.  Experimentally accessible quantities being independent of any free parameters are pointed out in the latter case.
\PACS{
      {21.60.Fw}{Models based on group theory}  \and
      {21.60.Ev}{Collective models}
     } 
} 
\authorrunning{ }
\titlerunning{ }
\maketitle
%

Proxy-SU(3) is an approximate symmetry appearing in heavy deformed nuclei \cite{proxy1,proxy2}. The foundations of proxy-SU(3) \cite{Assimakis}, its parameter-free predictions for the collective deformation
parameters $\beta$ and $\gamma$ \cite{Bonatsos,Martinou}, as well as for $B(E2)$ ratios \cite{Martinou}, have been discussed and its usefulness in explaining the dominance of prolate over oblate shapes in the ground states of even-even nuclei \cite{Sarantopoulou} and the point of the prolate to oblate shape transition in the rare earths region \cite{Sarantopoulou} has been demonstrated. In the present contribution, preliminary calculations for the spectra of heavy deformed nuclei, in which three-body and four-body operators are needed, will be discussed.  

Since Elliott demonstrated the relation of SU(3) symmetry to nuclear deformation \cite{Elliott1,Elliott2}, several group theoretical approaches to rotational nuclei have been developed. In theories approximating correlated valence nucleon pairs by bosons, like the Interacting Boson Model (IBM) \cite{IA}, the ground state band (gsb) is sitting in the lowest-lying irreducible representation (irrep) alone, while the $\gamma_1$ band and the $\beta_1$ band belong to the next irrep, therefore being degenerate to each other if only one-body and two-body terms are included in the Hamiltonian. Actually this degeneracy has been used as a hallmark of the appearance of SU(3) symmetry in atomic nuclei \cite{IA}. Higher order (three- and four-body terms) have been introduced in the IBM Hamiltonian mostly in order to accommodate triaxial shapes \cite{Chen,Moreau}. A particular class of higher order terms consists of the symmetry-preserving three-body operator $\Omega$ and the four-body operator $\Lambda$ (their mathematical names being the $O_l^0$ and $Q_l^0$ shift operator respectively) \cite{Hughes1,Hughes2,Judd,DeMeyer1}, the role of which in breaking the degeneracy between the $\beta_1$ and the $\gamma_1$ band \cite{DeMeyer2,Vanthournout}, as well as in producing the correct odd-even staggering within the $\gamma_1$ band \cite{BonPLB} has been considered. 

A different picture emerges within algebraic models employing fermions, like the pseudo-SU(3) \cite{pseudo1,pseudo2} and the proxy-SU(3) \cite{proxy1,proxy2} models. In these cases the lowest lying irrep accommodates both the gsb and the $\gamma_1$ band, and possibly higher-$K$ bands with $K=4$, 6, \dots, while the $\beta_1$ and $\gamma_2$ bands, and possibly  higher bands with $K=4$, 6, \dots belong to the next irrep. In these cases, the three- and/or four-body terms are absolutely necessary from the very beginning, in order to break the degeneracy between the gsb and the $\gamma_1$ bands. In the framework of pseudo-SU(3) this program has been succesfully carried out both by using general three- and four-body terms  \cite{DW1}, as well as by using a specific $K$-band splitting operator \cite{Naqvi}, containing the $\Omega$ and $\Lambda$ operators with appropriate coefficients. Numerical solutions have been produced in both cases, in the second case because the $\Lambda$ and $\Omega$ operators are diagonal in different bases \cite{DeMeyer2}. 

The $K$-band splitting operator used in \cite{Naqvi} has the interesting property of being diagonal for values of the angular momentum $L$ which are low in relation to the Elliott quantum numbers $\lambda$, $\mu$ characterizing the irreducible representations $(\lambda, \mu)$ of SU(3)  \cite{Elliott1,Elliott2}. In lowest order approximation, in what follows we are going to use the $K$ operator as a diagonal operator. 

In the present work we would like to consider the breaking of the degeneracy of the gsb and $\gamma_1$ band within the proxy-SU(3) scheme, using the same $\Lambda$, $\Omega$, and $K$-band splitting operators mentioned above. Before attempting any fittings, we would like to focus attention on physical quantities which exhibit  some characteristic behavior. For example, if we consider Hamiltonians of the form \cite{DeMeyer2}
\begin{equation}
H^{(3)} = a L^2 + b K + c \Omega - d L^4,  
\end{equation}  
or \begin{equation} \label{H2}
H^{(4)} = a L^2 + b K + c  \Lambda -d L^4,  
\end{equation} 
one can easily realize that the behavior of the differences of the energies of the gsb and the $\gamma_1$ bands for the same angular momentum $L$, $E(L_{\gamma_1})-E(L_g)$, normalized to their first member, $E(2_{\gamma_1})-E(2_g)$, will depend only on the relative parameter $c/b$, since only the second and the third term in the above Hamiltonians would contribute to them. Essentially parameter-independent predictions would also occur for the odd-even staggering \cite{stagg1,stagg2} within the $\gamma$-bands, which is essentially determined by the third term in the above Hamiltonians, while the first and fourth term have a minimal influence. It is interesting that while for the odd-even staggering detailed studies exist, pointing out the different behavior of this quantity in vibrational, rotational, $\gamma$-unstable or triaxial nuclei \cite{stagg1,stagg2}, no similar study exists for the behavior of the energy differences between the gsb and the $\gamma_1$ band in the different regions, thus we will first attempt such a study.

In Fig. 1 experimental values of $E(L_\gamma^+)-E(L_g^+)$ are plotted as a function of the angular momentum $L$ for several series of isotopes. For all isotopes normalization to $E(2_\gamma^+)-E(2_g^+)$ has been used. The following observations can be made.

1) In most of the deformed nuclei reported in these figures, the ``distance'' between the gsb and the $\gamma_1$ band is decreasing, the actinides been a clear example. 

2) Several examples of increasing ``distance'' are seen in the Os-Pt region, in which the O(6) symmetry is known to be present \cite{IA}.

3) Increasing ``distance'' is also seen in a few nuclei ($^{170}$Er, $^{192}$Os, $^{192}$Pt, which are expected to be triaxial, based on the staggering behavior exhibited by their $\gamma_1$ bands \cite{stagg2}. 

4) No effort has been made to exclude levels which are obviously due to band-crossing, like the last point shown in $^{188}$Pt. 

It should be noticed at this point, that the odd-even staggering in $\gamma_1$ bands, defined as 
\begin{equation}\label{stgg}
\Delta E (L) = E(L)- {E(L-1)+E(L+1) \over 2}, 
\end{equation}
 is also known to exhibit different behavior in various regions \cite{stagg1,stagg2}. In particular, staggering of small magnitude is seen in most of the deformed nuclei in the rare earths and in the actinides region, while strong staggering is seen in the Xe-Ba-Ce region. 

The present systematics of the energy differences between the gsb and the $\gamma_1$ band can be combined with the systematics of odd-even staggering in the $\gamma_1$-bands, which should be calculated and compared to the data. Since the sign in front of the three- or four-body term in the Hamiltonian has to be fixed in order to guarantee that the $\gamma_1$ band will lie above the gsb, the sign of the change of the ``distance'' between the $\gamma_1$ band and the gsb , as well as the form of the staggering within the  $\gamma_1$ bands (minima at even $L$ and maxima at odd $L$, or vice versa) are also be fixed 
by this choice, offering consistency checks of the symmetry. 

Preliminary proxy-SU(3) predictions for four deformed nuclei, obtained with the Hamiltonian of Eq. (\ref{H2}) with the parameters of Table 1, are shown in Fig. 2 for the ``distance'' between the $\gamma_1$ band and the gsb. In all cases decrease is predicted. Notice that the slope of the theoretical curve is determined by the parameter ratio $c/b$, while the parameter $b$ can be considered as a scale parameter for the energy differences under consideration. Parameters $a$ and $d$ do not influence these energy differences.

Results for the odd-even staggering within the $\gamma_1$ band for the same nuclei are shown in Fig. 3, in which the small energy scale should be noticed. In the results labeled ``2-terms'', only the second and the third terms of Eq. (\ref{H2}) are taken into account, in analogy to Fig. 2, while in the results labeled ``4-terms'' all four terms of Eq. (\ref{H2}) are considered. It is seen that the two extra terms have little effect on the staggering quantity and certainly do not affect its overall shape, exhibiting minima at even values of $L$ and maxima at odd values of $L$.

The spectra obtained for two of these nuclei are shown in Table 2. Details of the calculations will be given in a longer publication. 

In a series of papers \cite{Jolos1,Jolos2}, Jolos and von Brentano have shown, based on experimental data, that different mass coefficients should be used in the Bohr Hamiltonian for the ground state band and the $\gamma_1$ band. In order to show this, they use Grodzins products \cite{Grodzins} of excitation energies and B(E2) transition rates. The relation of the above findings to the work of Jolos and von Brentano should be considered in a next step, in which B(E2) transition rates will be included in the study. 

Financial support by the Bulgarian National Science Fund (BNSF) under Contract No. KP-06-N28/6 is gratefully acknowledged.


\begin{table} [htb]

\caption{Parameters (in units of keV) of the Hamiltonian of Eq. (\ref{H2}) for four nuclei. Data were taken from Ref. \cite{ENSDF}. $L_g$ ($L_\gamma$) denotes the maximum angular momentum for the ground state band ($\gamma$-band) included in the fit.  }

\begin{tabular}{rrrrrrr}
\hline\noalign{\smallskip}
nucleus &  $10^{-2}$ a  & b  &  $10^{-7}$ c  & $10^{-5}$ d & $L_g$ & $L_\gamma$ \\
\noalign{\smallskip}\hline\noalign{\smallskip}

$^{162}$Er & 1443 & 408 & 440 & 1258& 20 & 12\\
$^{160}$Dy & 1025 & 445 & 578 &  412& 28 & 23\\
$^{166}$Yb &  540 & 483 &2992 &  533& 24 & 13\\
$^{178}$Hf & 1225 & 588 & 748 &  890& 20 & 15\\
\noalign{\smallskip}\hline

\end{tabular}

\end{table}


\begin{table*} [htb]

\caption{Spectra of $^{162}$Er and $^{178}$Hf in keV, taken from Ref. \cite{ENSDF}, fitted by the Hamiltonian of Eq. (\ref{H2}). The parameter values used are given in Table 1. The rms deviations in keV are 34 and 58
 respectively. }

\begin{tabular}{rrrrr|rrrrr}
\hline  \noalign {\smallskip}
  &$^{162}$Er & $^{162}$Er & $^{178}$Hf & $^{178}$Hf & 
  & $^{162}$Er & $^{162}$Er & $^{178}$Hf & $^{178}$Hf \\
  & exp & th & exp & th  &  & exp & th & exp & th \\ 
$L$     & & & & &$L$ & & & & \\
\noalign{\smallskip}\hline\noalign{\smallskip}

 2&  102&   93&   93&   85& 2&  901&  895& 1175& 1198\\
 4&  330&  309&  307&  291& 3& 1002&  987& 1269& 1282\\
 6&  667&  640&  632&  615& 4& 1128& 1107& 1384& 1372\\
 8& 1097& 1073& 1059& 1044& 5& 1286& 1259& 1533& 1529\\
10& 1603& 1588& 1570& 1566& 6& 1460& 1428& 1691& 1651\\
12& 2165& 2162& 2150& 2161& 7& 1669& 1638& 1890& 1876\\
14& 2746& 2766& 2777& 2809& 8& 1873& 1846& 2082& 2038\\
16& 3292& 3364& 3435& 3484& 9& 2134& 2107& 2316& 2295\\
18& 3847& 3920& 4119& 4157&10& 2347& 2344& 2538& 2519\\
20& 4463& 4388& 4837& 4795&11& 2656& 2647& 2798& 2782\\
22&     & 4719&     & 5361&12& 2911& 2901& 3053& 3073\\
24&     & 4859&     & 5814&13&     & 3224& 3336& 3316\\
26&     &     &     &     &14&     & 3488& 3625& 3680\\
28&     &     &     &     &15&     & 3810& 3928& 3874\\
\noalign{\smallskip}\hline
  
\end{tabular}
\end{table*}


\begin{figure*}[htb]

{\includegraphics[width=70mm]{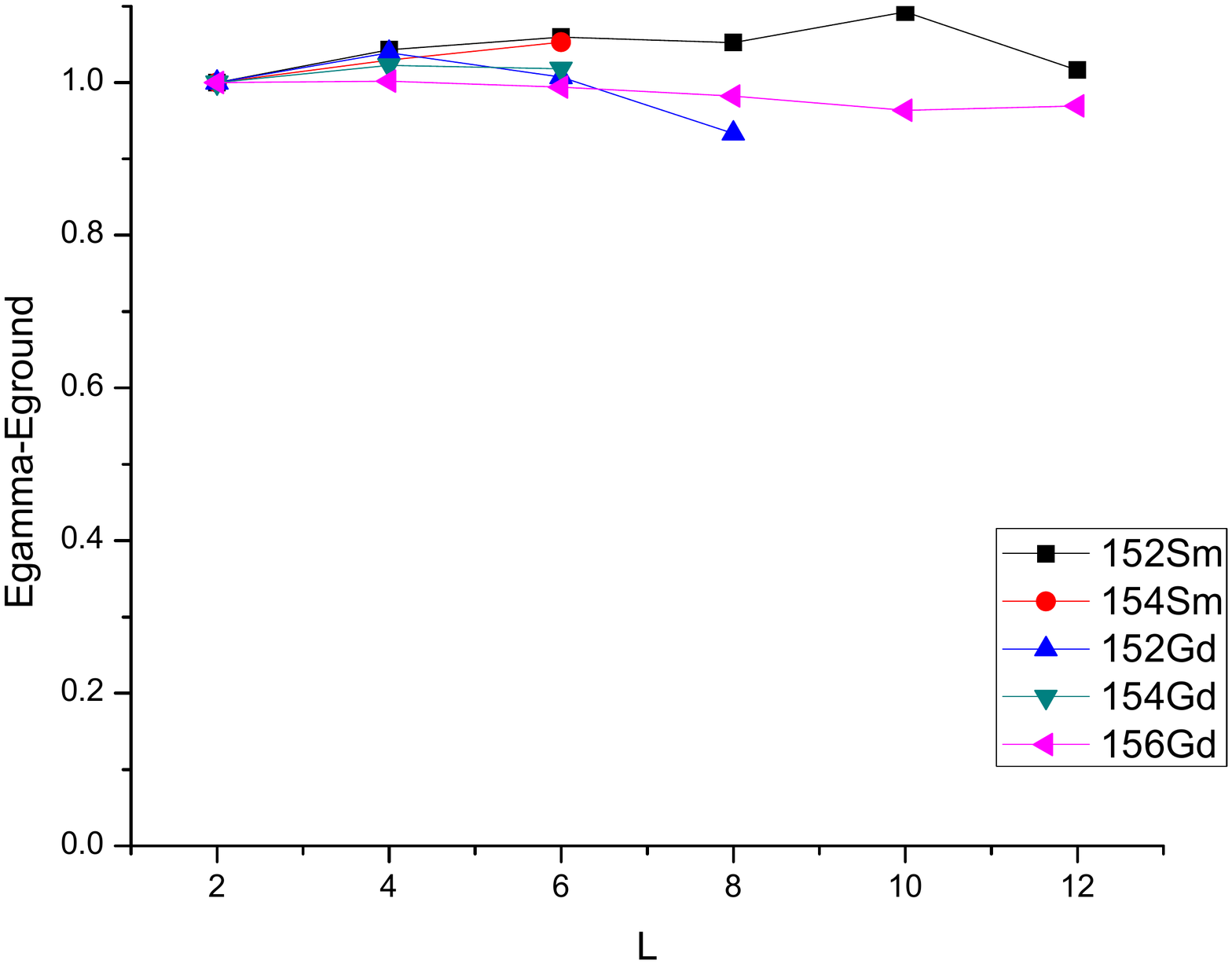}
\includegraphics[width=70mm]{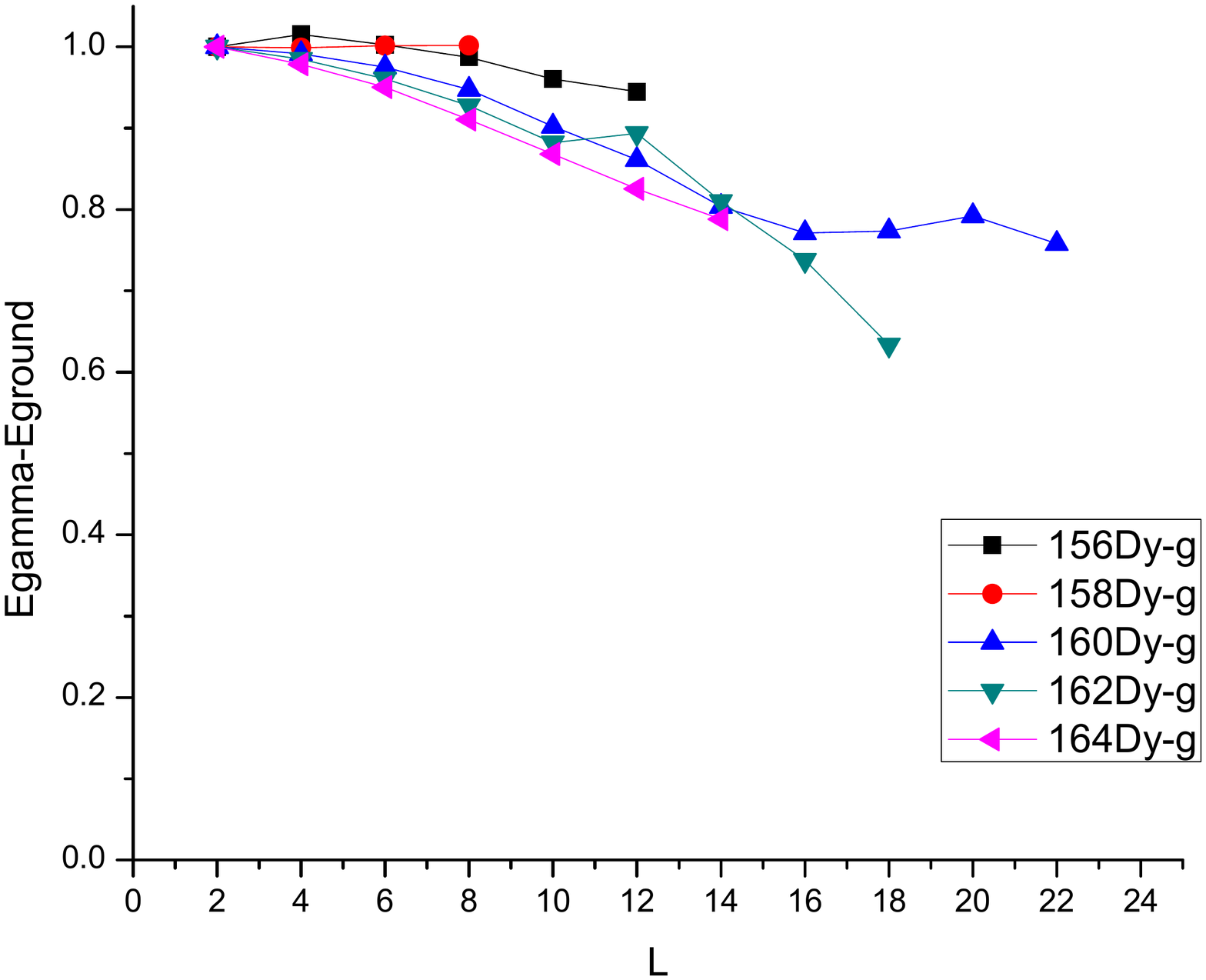}}

{\includegraphics[width=70mm]{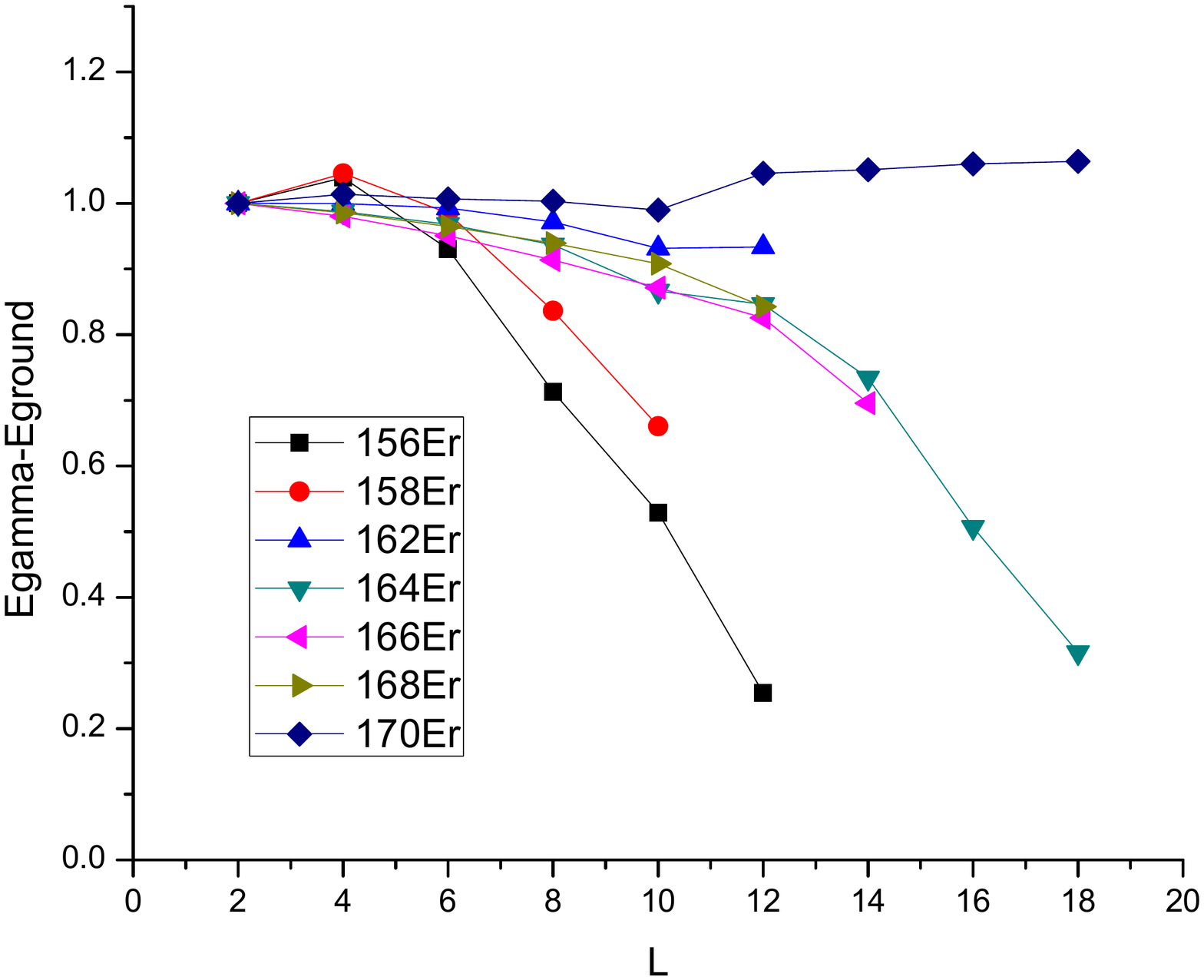}
\includegraphics[width=70mm]{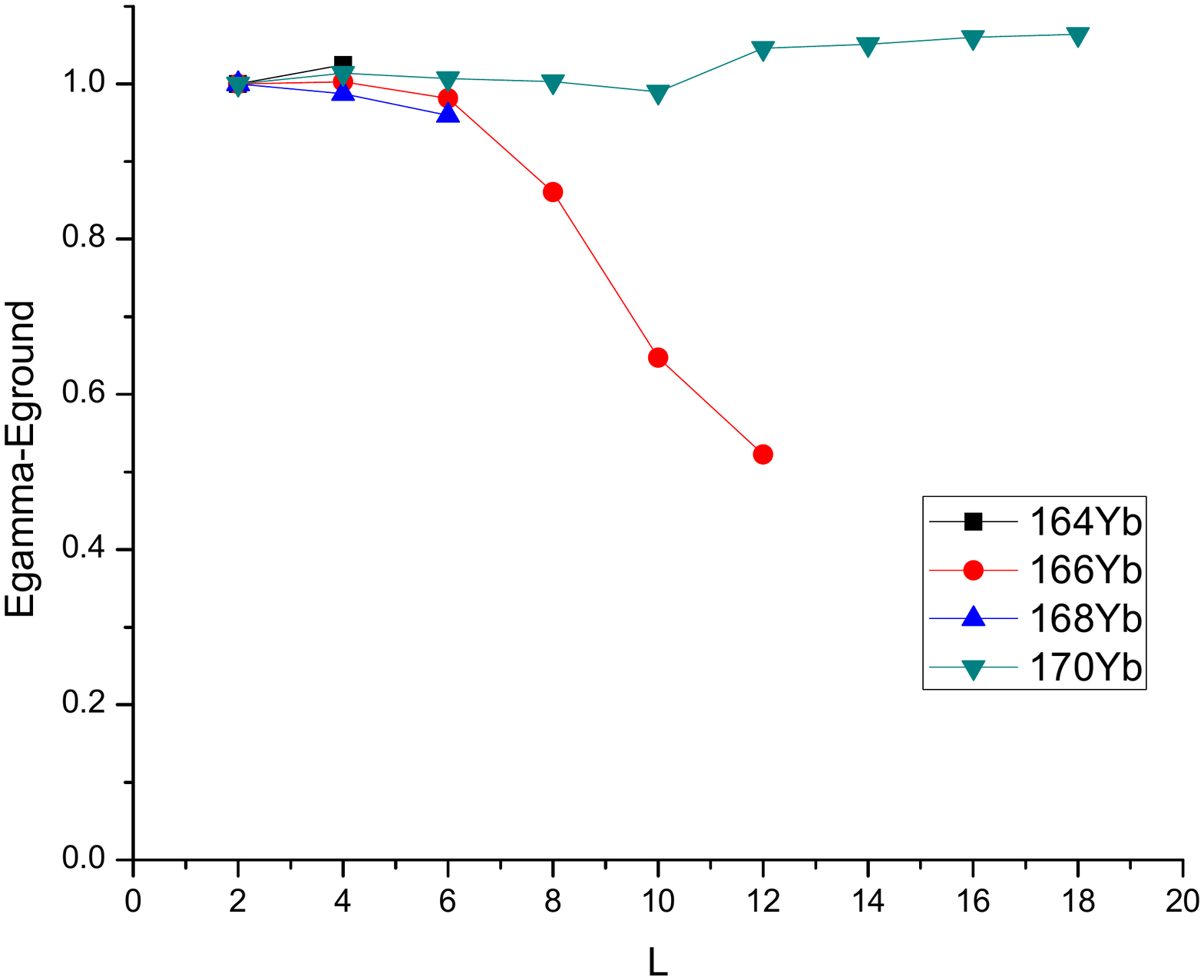}}

{\includegraphics[width=70mm]{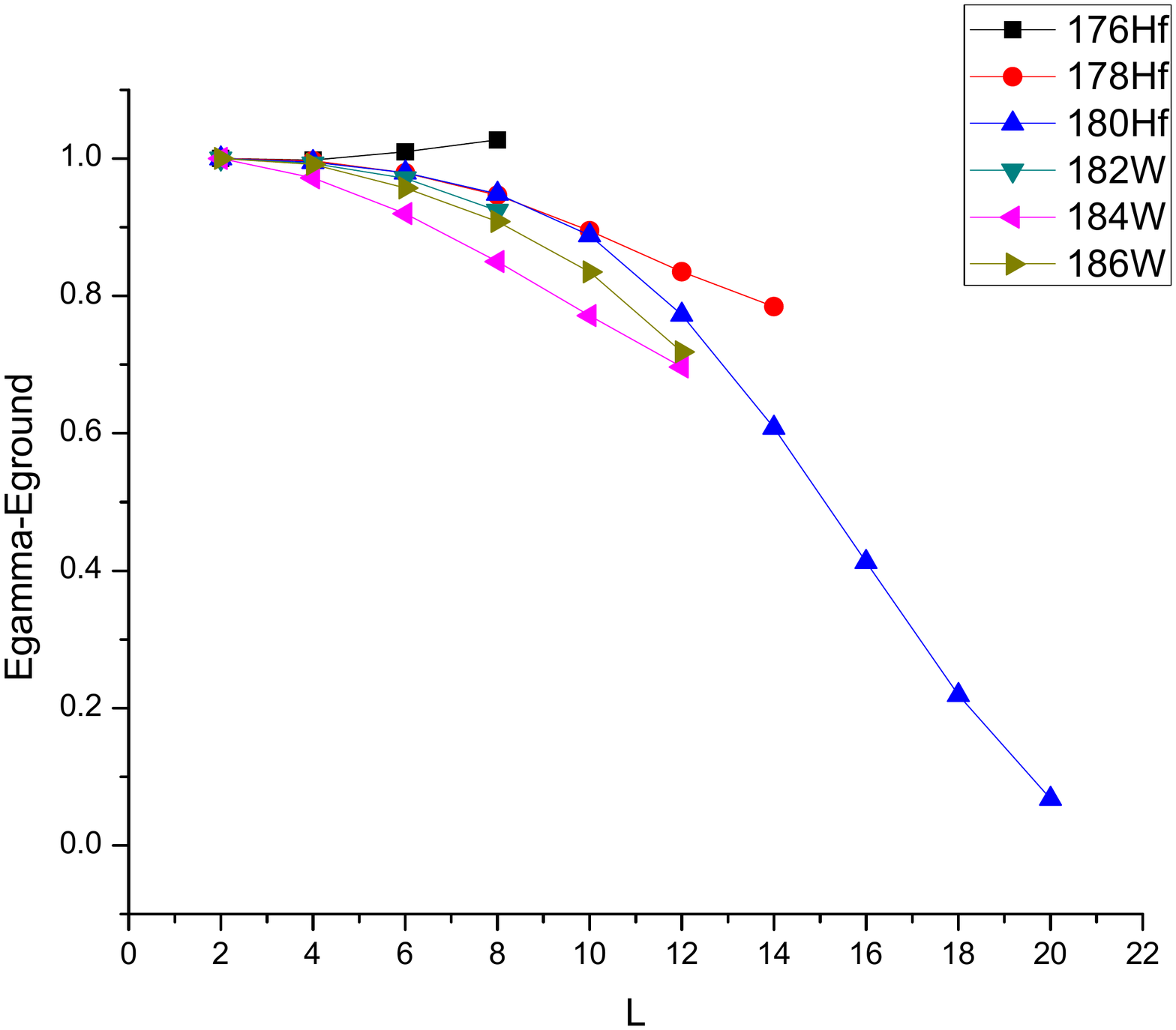}
\includegraphics[width=70mm]{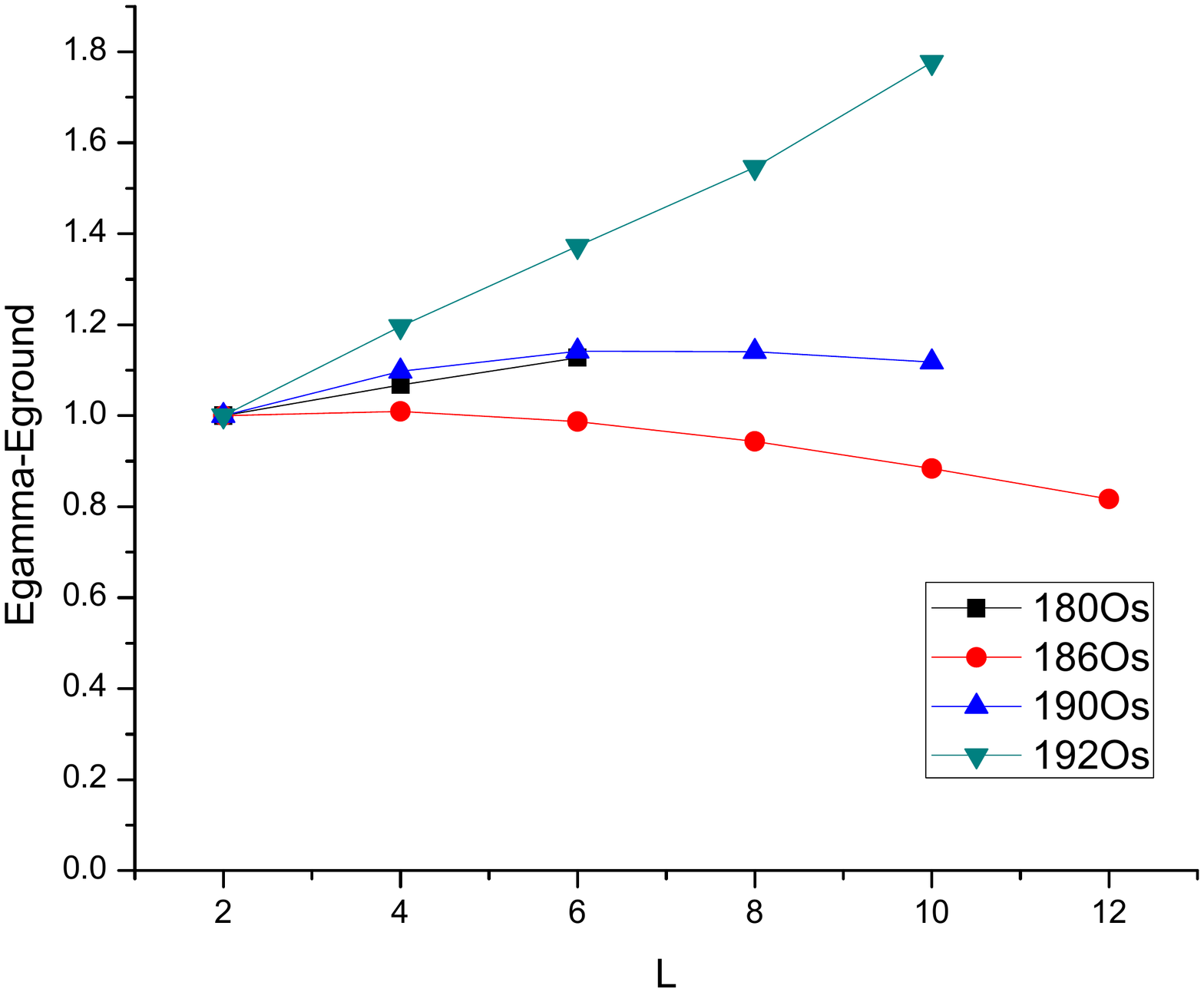}}

{\includegraphics[width=70mm]{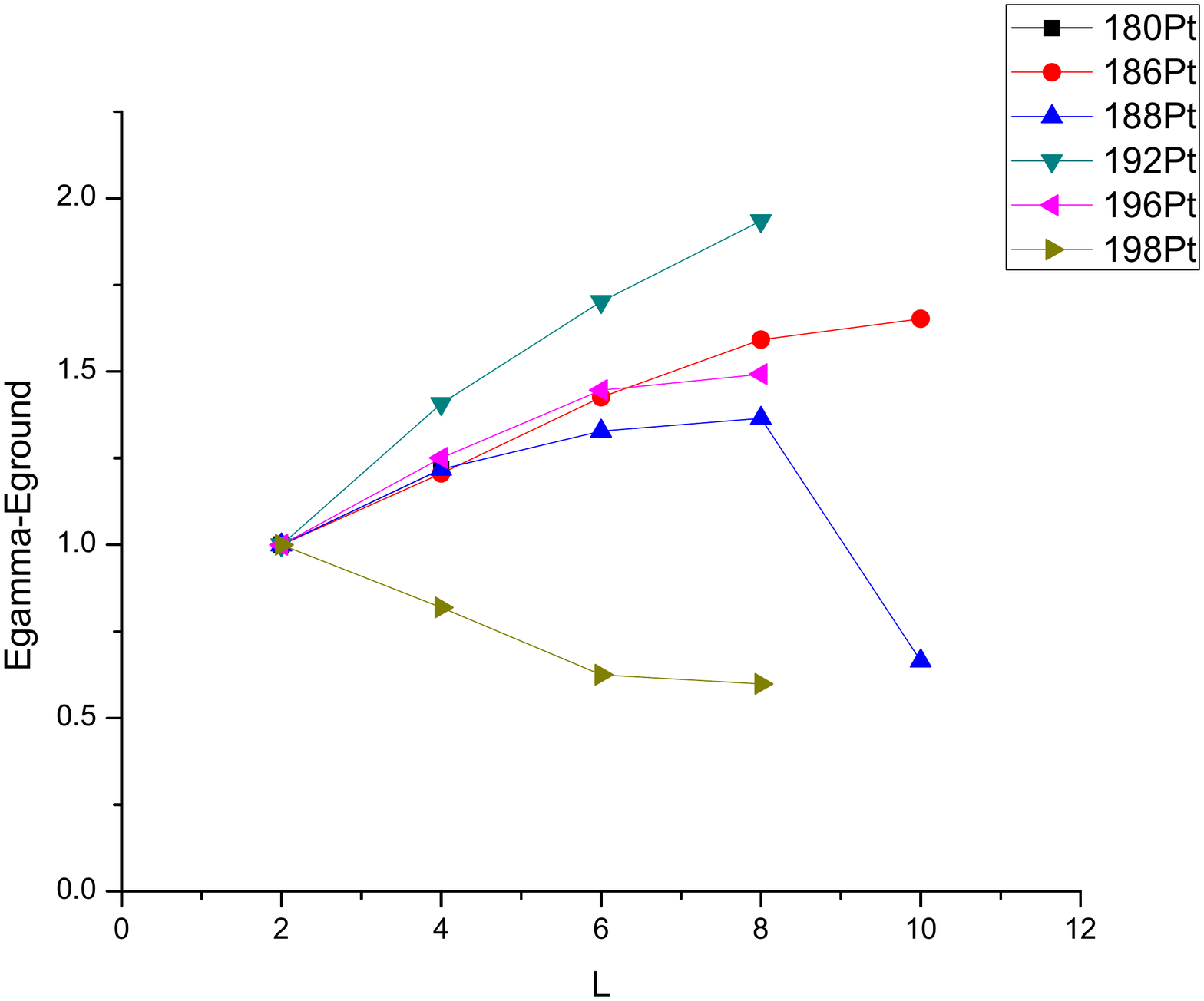}
\includegraphics[width=70mm]{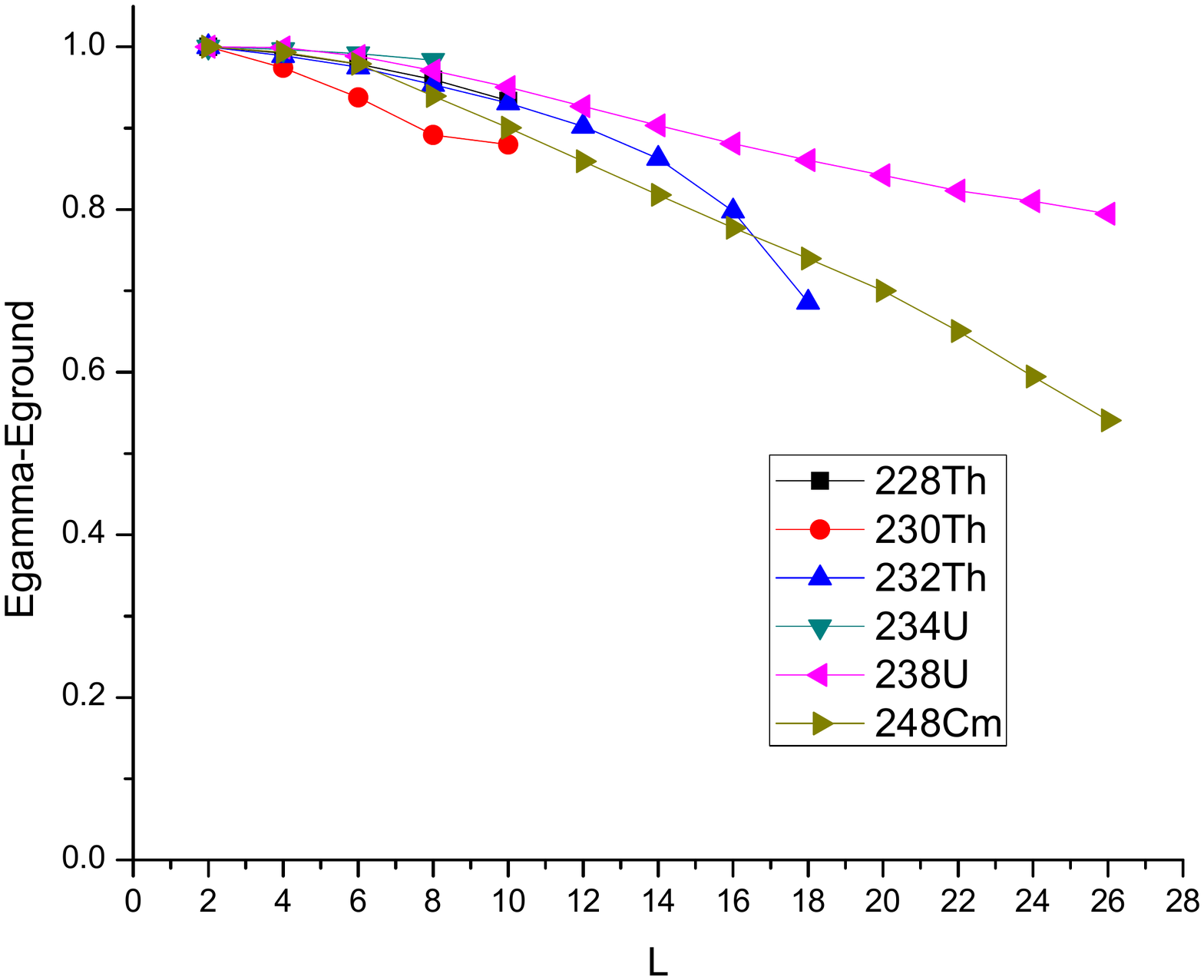}}

\caption{Experimental values of $E(L_\gamma)-E(L_g)$, taken from Ref. \cite{ENSDF}, plotted as function of the angular momentum $L$ for several series of isotopes. For all isotopes, normalization to $E(2_\gamma)-E(2_g)$ has been used.  
} 

\end{figure*}


\begin{figure*}[htb]
{\includegraphics[width=65mm]{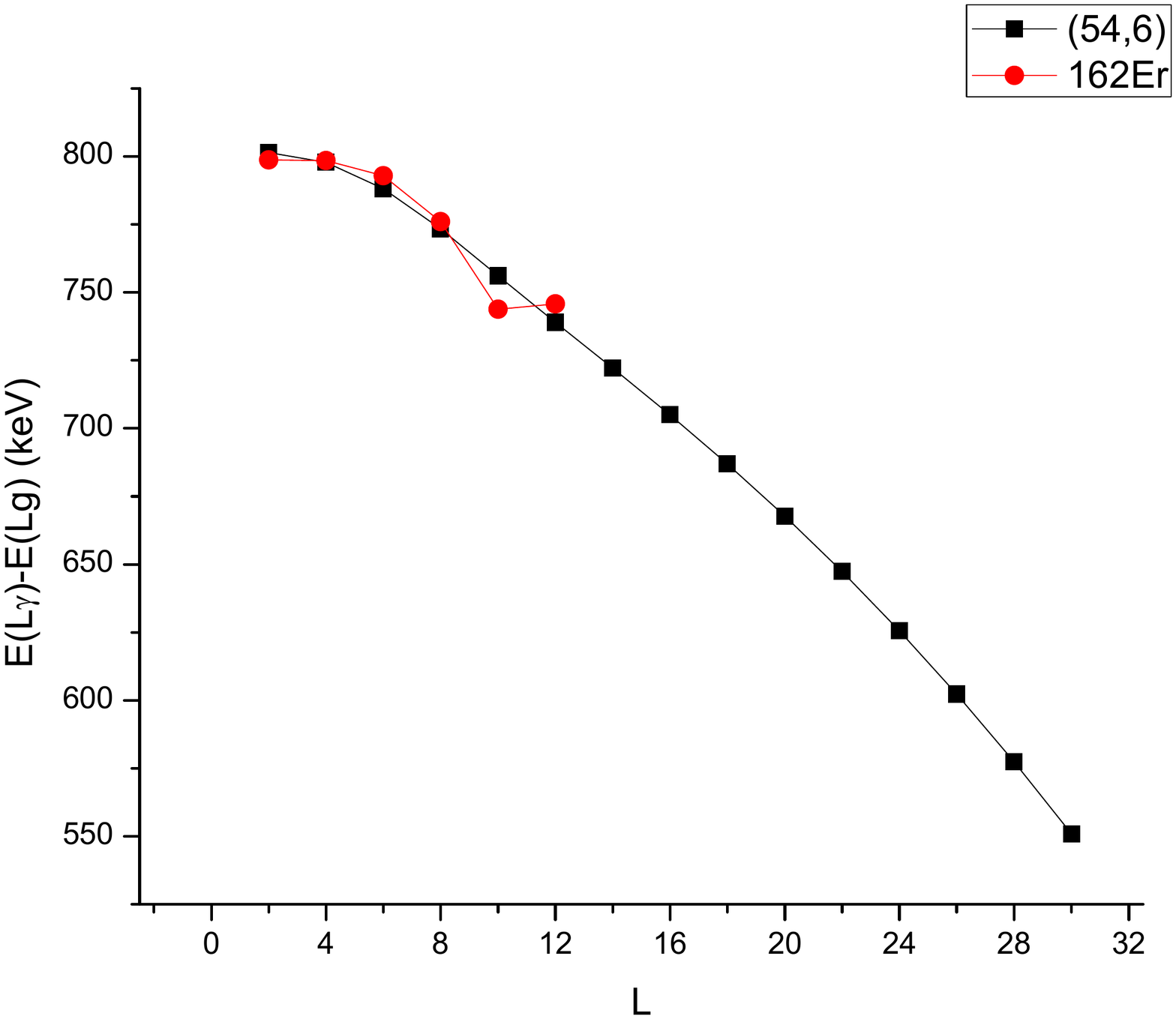}
\includegraphics[width=65mm]{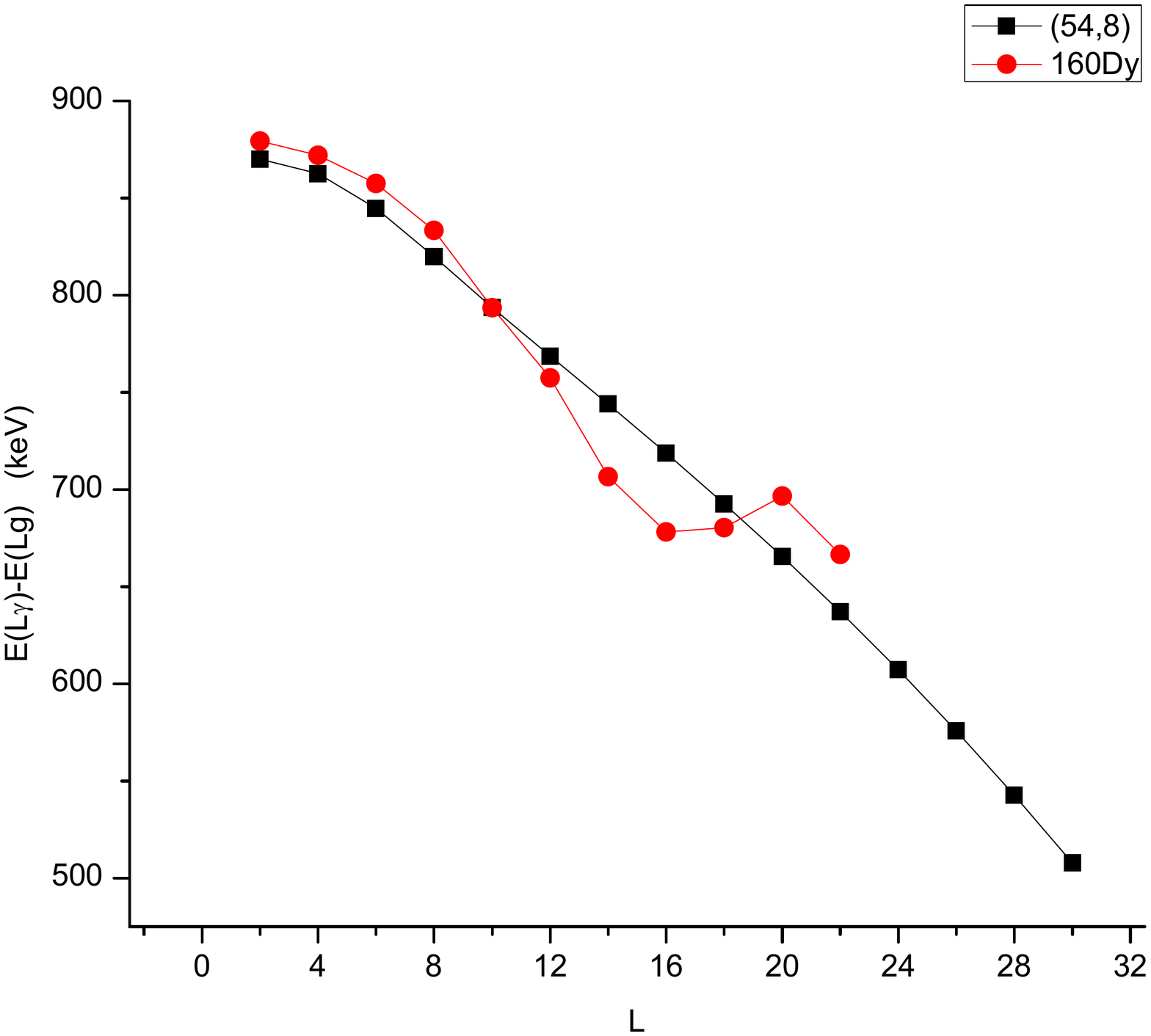}}

{\includegraphics[width=65mm]{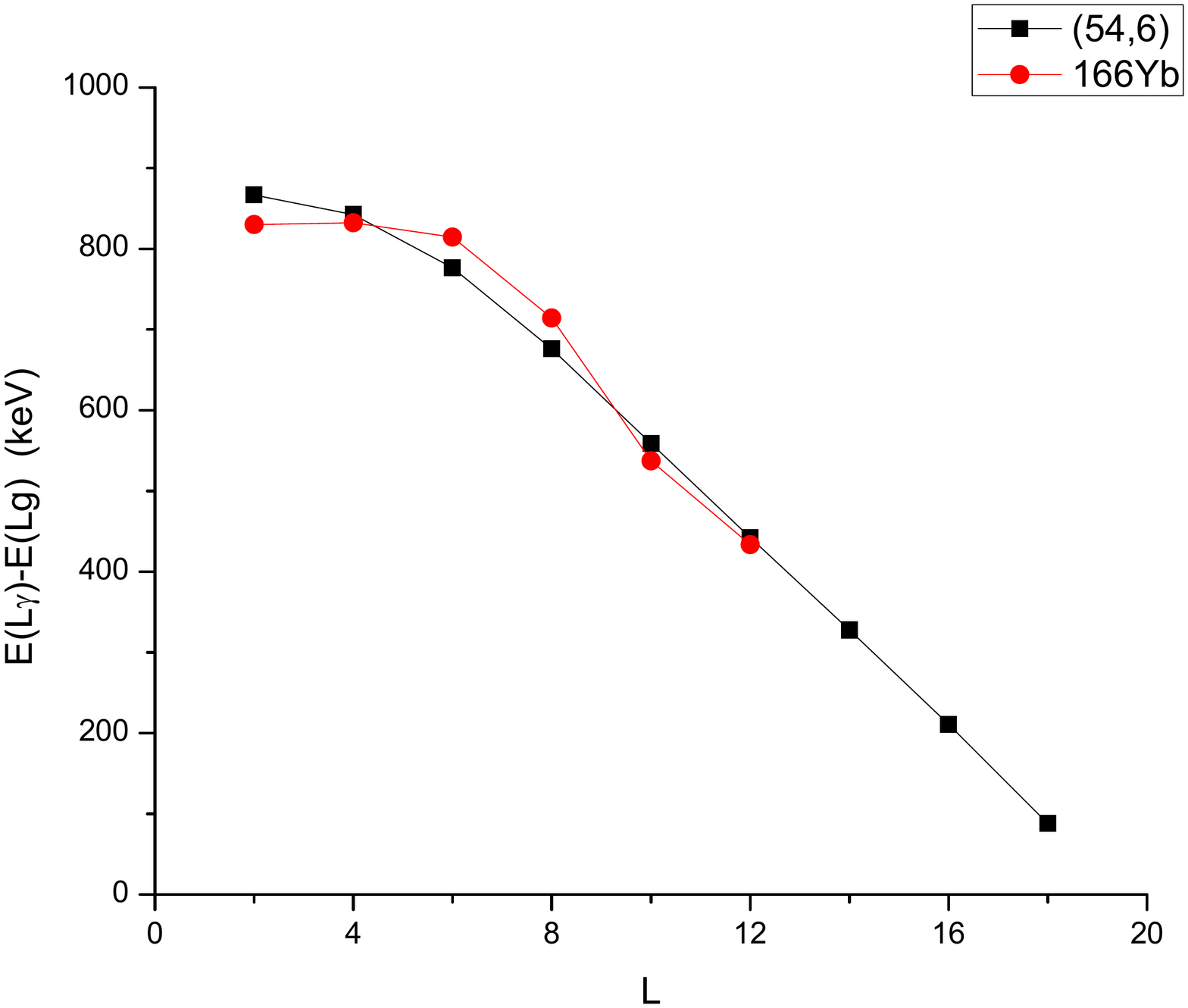}
\includegraphics[width=65mm]{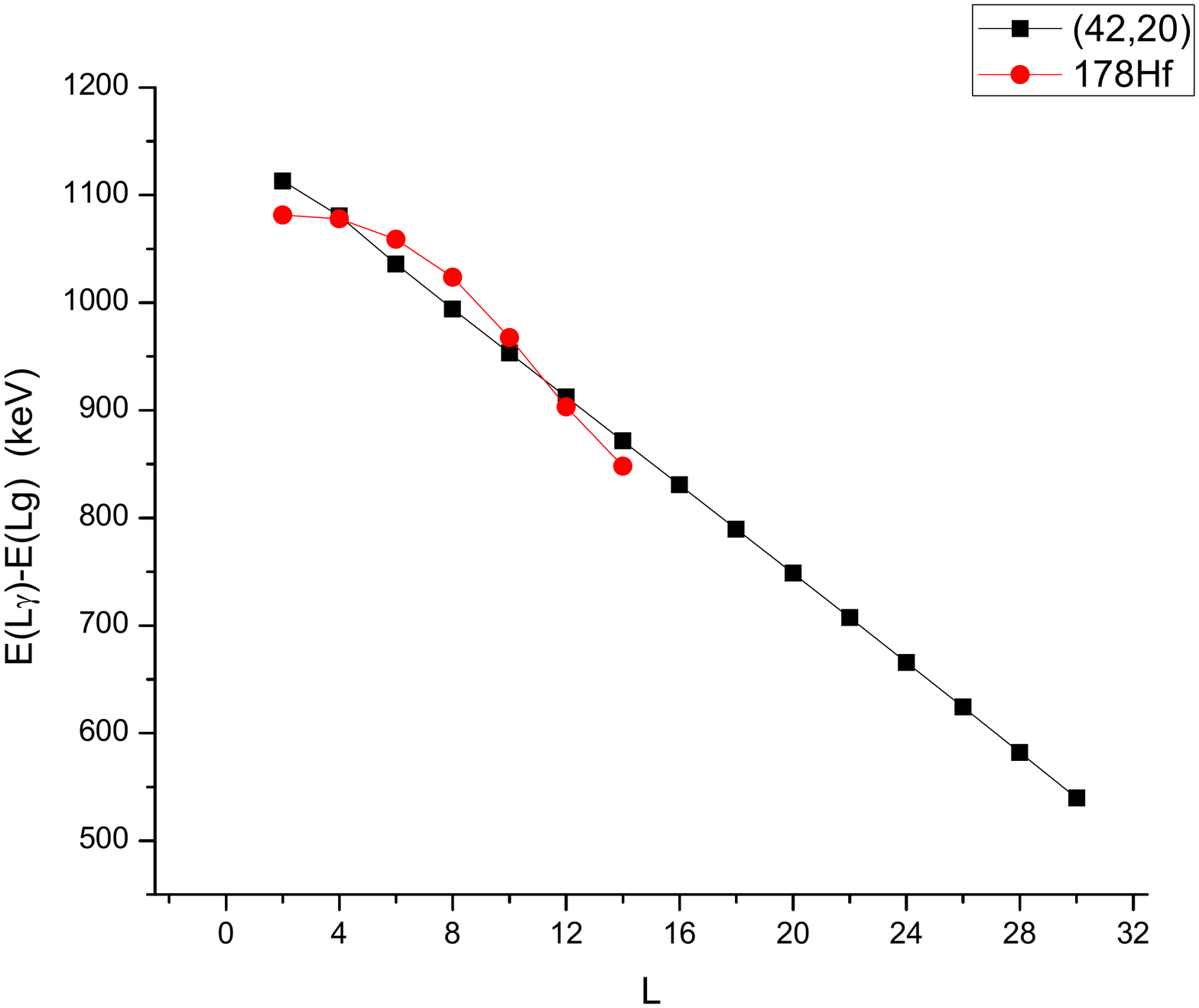}}

\caption{Experimental values of $E(L_\gamma)-E(L_g)$ \cite{ENSDF} compared to  proxy-SU(3) predictions from the Hamiltonian of Eq. (\ref{H2}) for four nuclei.
}

\end{figure*}


\begin{figure*}[htb]
{\includegraphics[width=65mm]{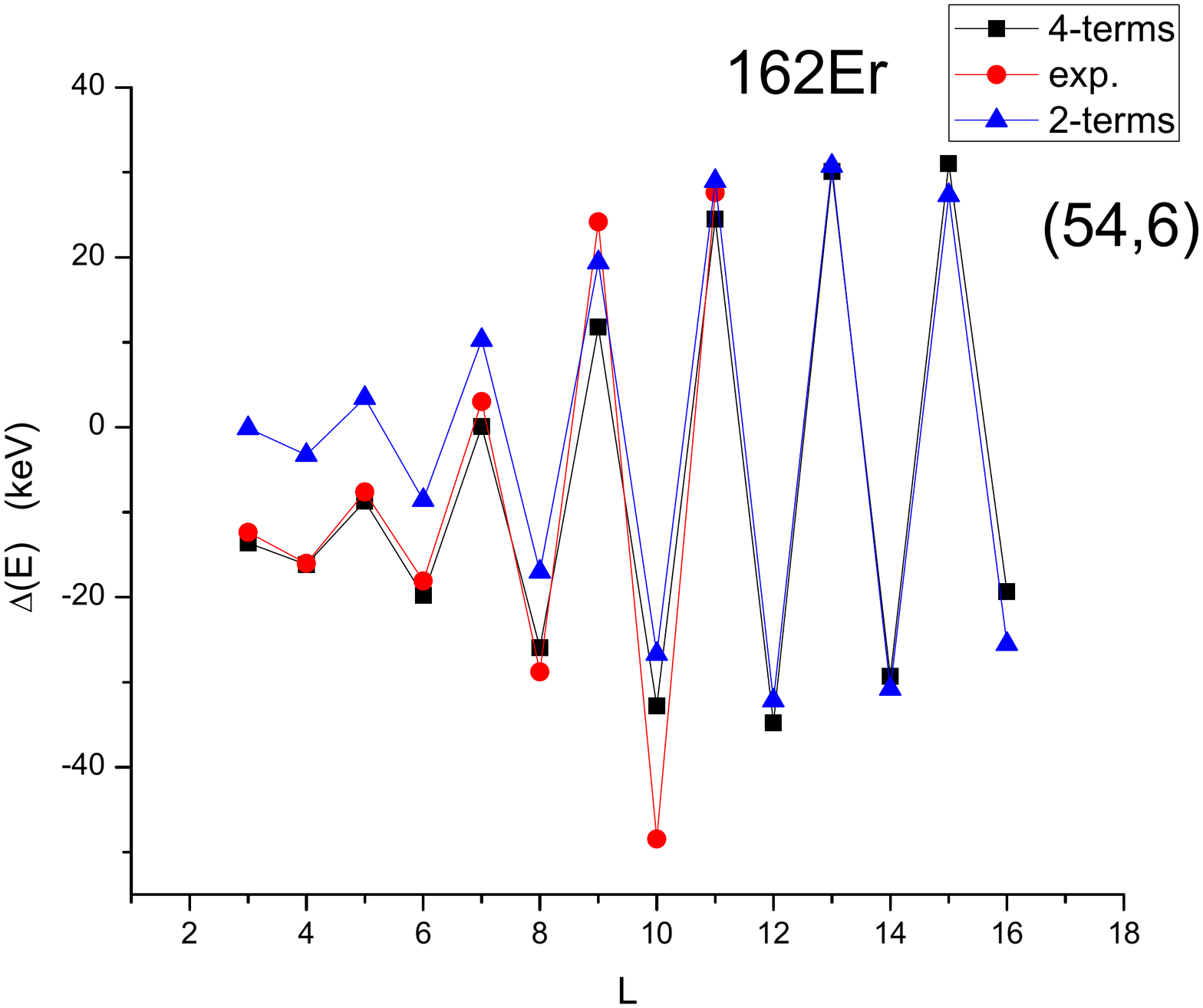}
\includegraphics[width=65mm]{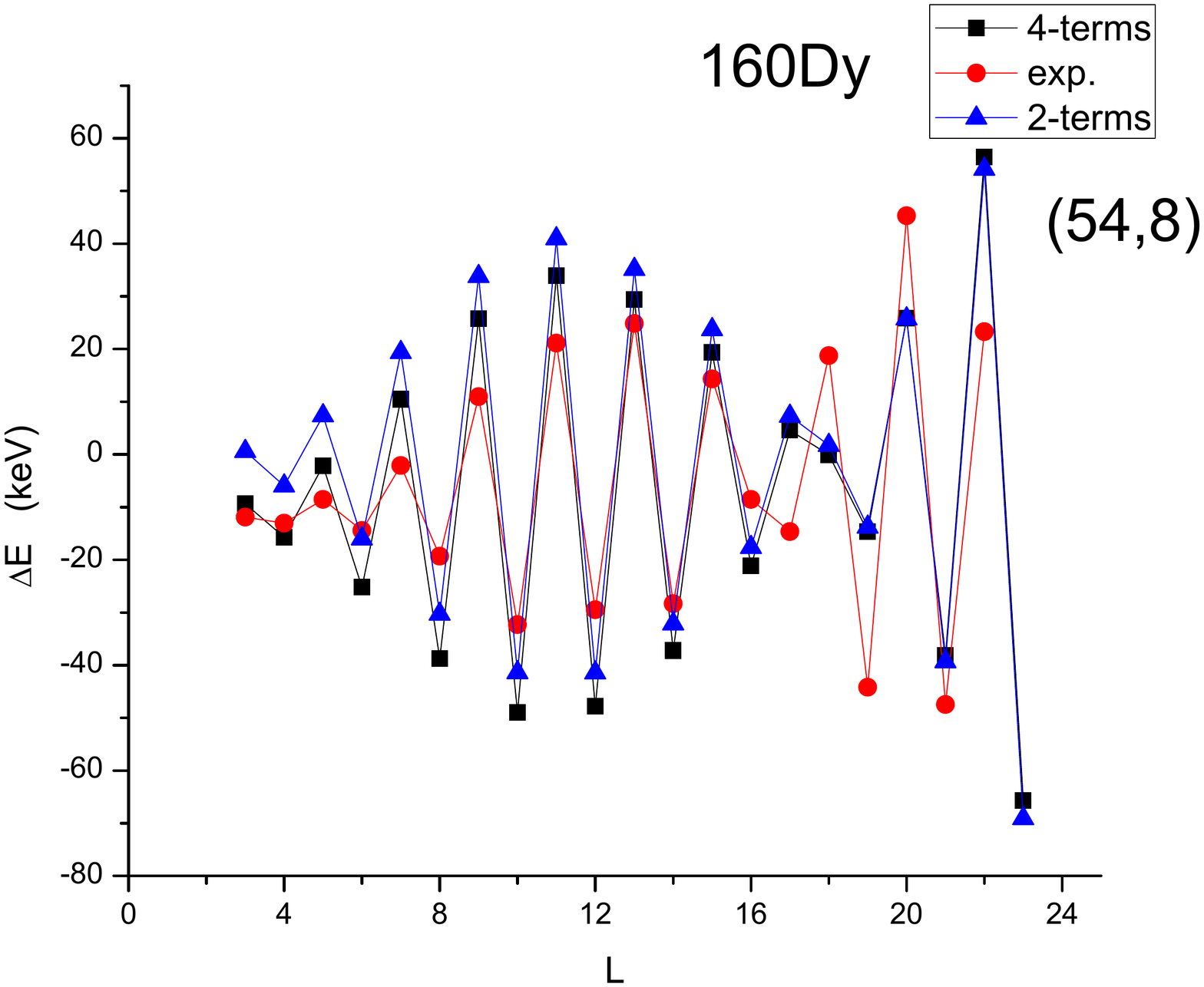}}

{\includegraphics[width=65mm]{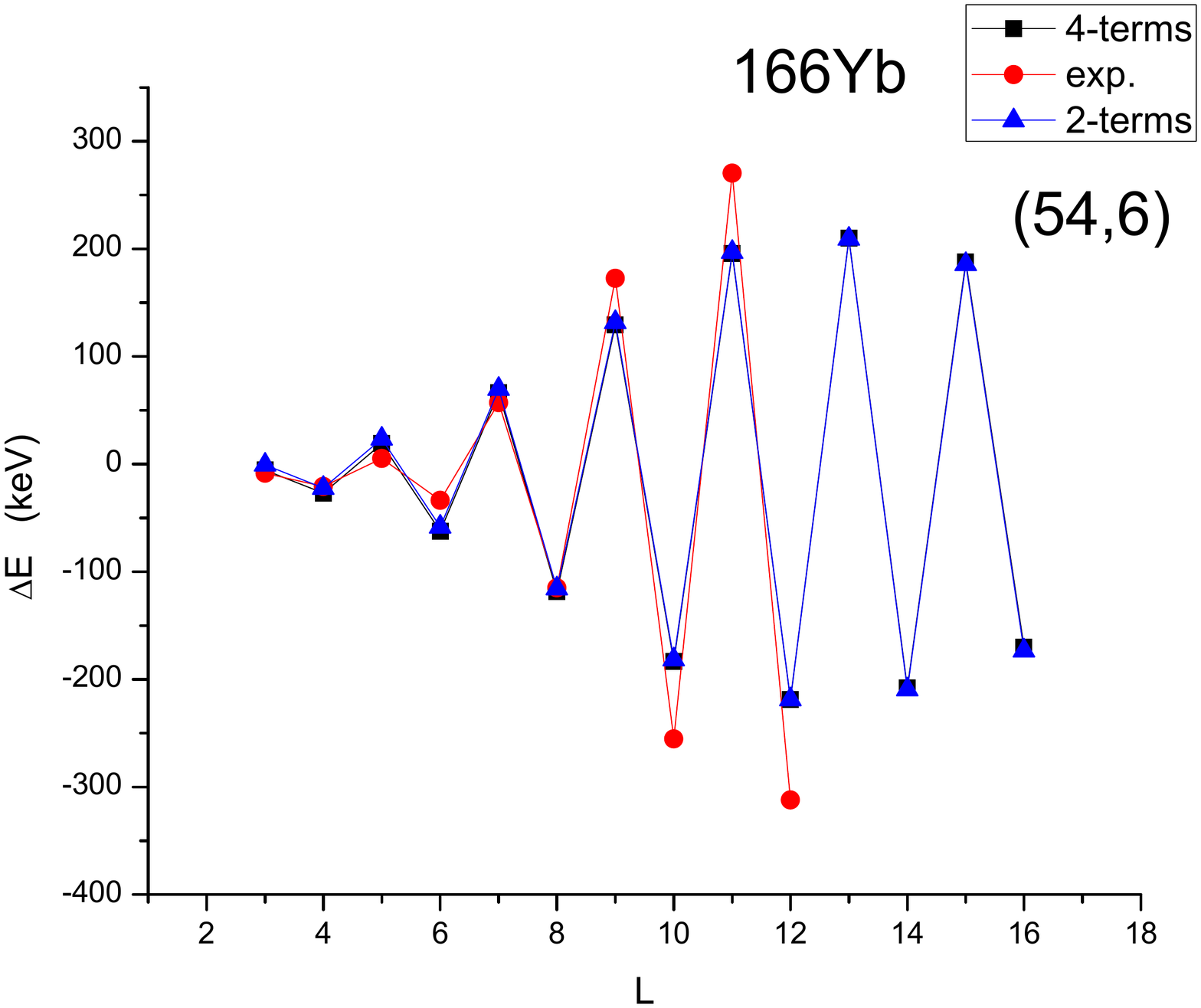}
\includegraphics[width=65mm]{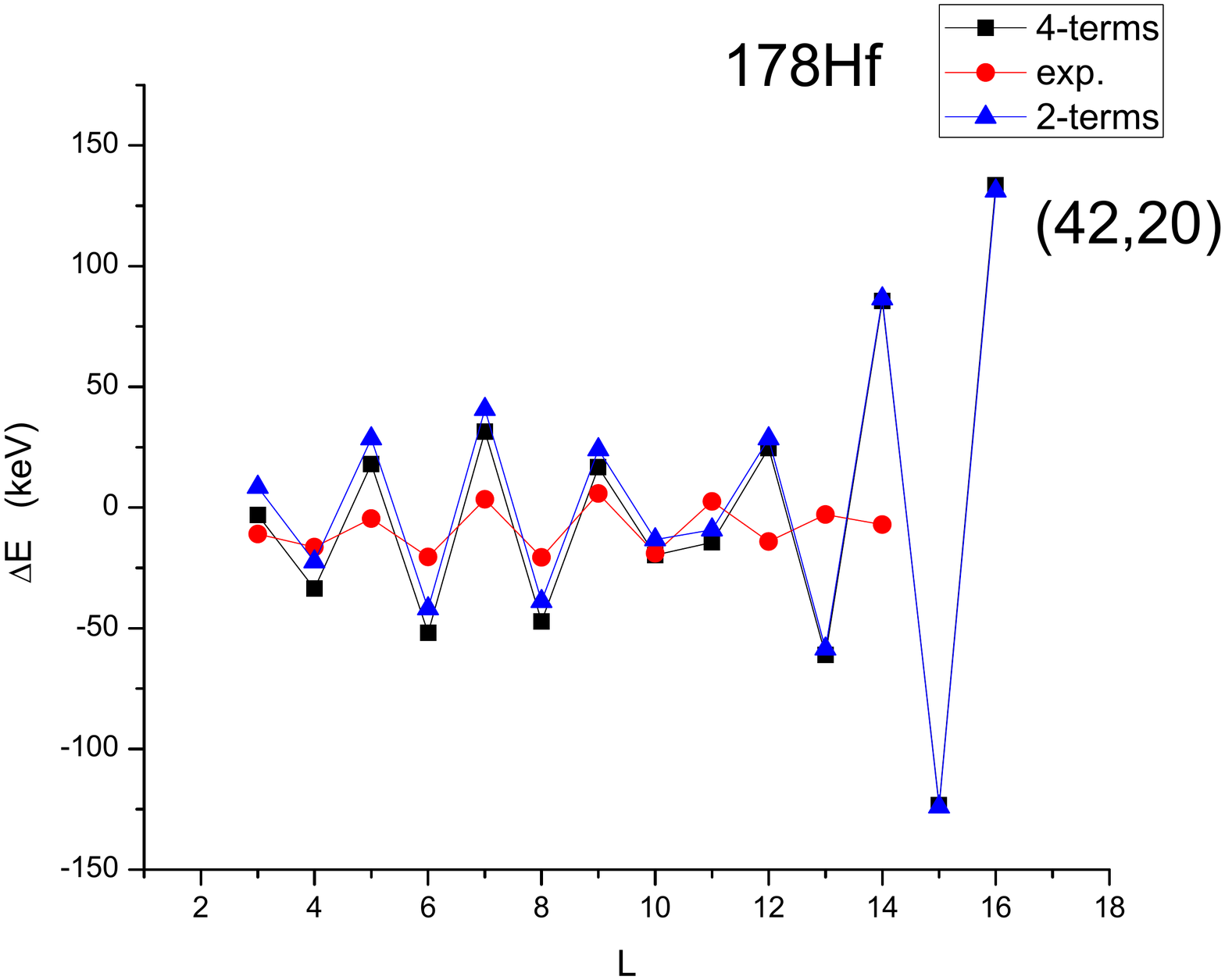}}

\caption{Experimental values of odd-even staggering in the $\gamma_1$ bands, calculated from Eq. (\ref{stgg}) using data from \cite{ENSDF}, compared  to proxy-SU(3) predictions from the Hamiltonian of Eq. (\ref{H2}) for four nuclei.
} 

\end{figure*}


\begin{thebibliography}{99}

\bibitem{proxy1} D. Bonatsos, I.E. Assimakis, N. Minkov, A. Martinou, R.B. Cakirli, R.F. Casten, and K. Blaum, Phys. Rev. C \textbf{95}, 064325 (2017)

\bibitem{proxy2} D. Bonatsos, I.E. Assimakis, N. Minkov, A. Martinou, S. Sarantopoulou, R.B. Cakirli, R.F. Casten, and K. Blaum, Phys. Rev. C \textbf{95}, 064326 (2017)

\bibitem{Assimakis} I.E. Assimakis,  D. Bonatsos,  N. Minkov, A. Martinou, R.B. Cakirli, R.F. Casten, and K. Blaum, Bulg. J. Phys. \textbf{44}, 398 (2017) 

\bibitem{Bonatsos} D. Bonatsos, I.E. Assimakis, N. Minkov, A. Martinou, S.K. Peroulis, S. Sarantopoulou,   
 R.B. Cakirli, R.F. Casten, and K. Blaum, Bulg. J. Phys. \textbf{44}, 385 (2017)

\bibitem{Martinou} A. Martinou, D. Bonatsos, I.E. Assimakis, N. Minkov,  S. Sarantopoulou, R.B. Cakirli, R.F. Casten, and K. Blaum, Bulg. J. Phys. \textbf{44}, 407  (2017)  

\bibitem{Sarantopoulou} S. Sarantopoulou, D. Bonatsos, I.E. Assimakis, N. Minkov, A. Martinou, R.B. Cakirli, R.F. Casten, and K. Blaum, Bulg. J. Phys. \textbf{44}, 417 (2017)  

\bibitem{Elliott1}
J. P. Elliott, Proc. Roy. Soc. Ser. A \textbf{245}, 128 (1958) 

\bibitem{Elliott2}
J. P. Elliott, Proc. Roy. Soc. Ser. A \textbf{245}, 562 (1958)

\bibitem{IA}
F. Iachello and A. Arima, {\it The Interacting Boson Model} (Cambridge University Press, Cambridge, 1987)

\bibitem{Chen}
P. Van Isacker and J.-Q. Chen, Phys. Rev. C \textbf{24}, 684 (1981) 

\bibitem{Moreau} K. Heyde, P. Van Isacker, M. Waroquier, and J. Moreau, Phys. Rev. C \textbf{29}, 1420 (1984) 

\bibitem{Hughes1}
J.W.B. Hughes, J. Phys. A: Math., Nucl. Gen. \textbf{6}, 48 (1973) 

\bibitem{Hughes2}
J.W.B. Hughes, J. Phys. A: Math., Nucl. Gen. \textbf{6}, 281 (1973) 

\bibitem{Judd}
B.R. Judd, W. Miller, Jr., J. Patera, and P. Winternitz, J. Math. Phys. \textbf{15}, 1787 (1974) 

\bibitem{DeMeyer1}
H. De Meyer, G. Vanden Berghe, and J. Van der Jeugt, J. Math. Phys. \textbf{26}, 3109 (1985) 

\bibitem{DeMeyer2}
G. Vanden Berghe, H.E. De Meyer, and P. Van Isacker, Phys. Rev. C \textbf{32}, 1049 (1985) 

\bibitem{Vanthournout}
J. Vanthournout, Phys. Rev. C \textbf{41}, 2380 (1990) 

\bibitem{BonPLB}
D. Bonatsos, Phys. Lett. B \textbf{200}, 1 (1988) 

\bibitem{pseudo1}
R. D. Ratna Raju, J. P. Draayer, and K. T. Hecht, Nucl. Phys. A \textbf{202}, 433 (1973)

\bibitem{pseudo2}
J. P. Draayer, K. J. Weeks, and K. T. Hecht,  Nucl. Phys. A \textbf{381}, 1 (1982) 

\bibitem{DW1}
J. P. Draayer and K. J. Weeks,  Ann. Phys. (N.Y.) \textbf{156}, 41 (1984)

\bibitem{Naqvi}
H.A. Naqvi and J.P. Draayer, Nucl. Phys. A  \textbf{516}, 351 (1990) 

\bibitem{stagg1}
 N.V. Zamfir, and R.F. Casten, Phys. Lett. B \textbf{260}, 265 (1991) 

\bibitem{stagg2}
E.A. McCutchan, D. Bonatsos, N.V. Zamfir, and R.F. Casten, Phys. Rev. C \textbf{76}, 024306 (2007) 

\bibitem{ENSDF}
Brookhaven National Laboratory ENSDF database http://www.nndc.bnl.gov/ensdf/

\bibitem{Jolos1} 
R. V. Jolos and P. von Brentano,  Phys. Rev. C \textbf{74}, 064307 (2006)

\bibitem{Jolos2} 
R. V. Jolos and P. von Brentano,  Phys. Rev. C \textbf{76}, 024309 (2007)

\bibitem{Grodzins}
L. Grodzins, Phys. Lett. \textbf{2}, 88 (1962) 

\end{thebibliography}
\end{document}